\documentclass[10pt,conference]{IEEEtran} 
\IEEEoverridecommandlockouts
\usepackage{cite}
\usepackage{amsmath,amssymb,amsfonts}
\usepackage{algorithm}
\usepackage{algorithmic}
\usepackage{graphicx}
\usepackage{cuted}
\graphicspath{{fig/}}
\usepackage{textcomp}
\usepackage{xcolor}
\definecolor{codegreen}{rgb}{0,0.5,0}
\definecolor{codeblue}{rgb}{0,0,0.7}
\definecolor{codegray}{rgb}{0.45,0.45,0.45}
\definecolor{codeteal}{rgb}{0,0.4,0.4}
\definecolor{coderule}{rgb}{0.75,0.75,0.75}
\definecolor{answerbg}{RGB}{233,240,248}
\usepackage{listings}
\usepackage{caption}
\usepackage{booktabs}
\usepackage{multirow}
\usepackage{pgfplots}
\usepackage{siunitx}
\usepackage{threeparttable}
\pgfplotsset{compat=1.18}

\lstdefinelanguage{Rust}{
  keywords={as, async, await, break, const, continue, crate, dyn, else, enum,
    extern, false, fn, for, if, impl, in, let, loop, match, mod, move, mut,
    pub, ref, return, self, Self, static, struct, super, trait, true, type,
    unsafe, use, where, while, yield},
  sensitive=true,
  morecomment=[l]{//},
  morecomment=[s]{/*}{*/},
  morestring=[b]{"},
  morestring=[b]{'},
}

\def\BibTeX{{\rm B\kern-.05em{\sc i\kern-.025em b}\kern-.08em
    T\kern-.1667em\lower.7ex\hbox{E}\kern-.125emX}}

\newcommand{\best}[1]{\textbf{#1}}
\newcommand{\tied}[1]{\underline{#1}}
\newcommand{\rqanswer}[2]{%
  \par\smallskip
  \noindent\colorbox{answerbg}{%
    \parbox{\dimexpr\columnwidth-2\fboxsep\relax}{%
      \textbf{Answer to #1:} #2%
    }%
  }%
  \par\smallskip
}
\newcommand{\codebad}[1]{%
  \tikz[baseline=(code.base)]{
    \node[
      anchor=base,
      outer sep=0pt,
      inner xsep=0.8pt,
      inner ysep=0.35pt,
      rounded corners=1.2pt,
      fill=red!16,
      text height=1.8ex,
      text depth=0.4ex
    ] (code) {\ttfamily\tiny #1};
  }%
}
\newcommand{\codegood}[1]{%
  \tikz[baseline=(code.base)]{
    \node[
      anchor=base,
      outer sep=0pt,
      inner xsep=0.8pt,
      inner ysep=0.35pt,
      rounded corners=1.2pt,
      fill=cyan!18,
      text height=1.8ex,
      text depth=0.4ex
    ] (code) {\ttfamily\tiny #1};
  }%
}
\begin{document}

\title{SpeC2Rust: Specification-Driven Repository-Level C-to-Rust Translation}
% \author{Anonymous Author(s)}

\makeatletter
\newcommand{\linebreakand}{%
  \end{@IEEEauthorhalign}
  \hfill\mbox{}\par
  \vspace{1em}%
  \mbox{}\hfill\begin{@IEEEauthorhalign}
}
\makeatother

\author{%
\IEEEauthorblockN{Junjie Yao\textsuperscript{*}}
\IEEEauthorblockA{%
\textit{Harbin Institute of Technology}\\
\textit{Shenzhen, China}\\
220110716@stu.hit.edu.cn}
\and
\IEEEauthorblockN{Xiongzheng Guan\textsuperscript{*}}
\IEEEauthorblockA{%
\textit{Harbin Institute of Technology}\\
\textit{Shenzhen, China}\\
nancicv@163.com}
\and
\IEEEauthorblockN{Jia Feng}
\IEEEauthorblockA{%
\textit{Harbin Institute of Technology}\\
\textit{Shenzhen, China}\\
jiafeng@stu.hit.edu.cn}
\linebreakand
\IEEEauthorblockN{Yujia Chen}
\IEEEauthorblockA{%
\textit{Harbin Institute of Technology}\\
\textit{Shenzhen, China}\\
yujiachen@stu.hit.edu.cn}
\and
\IEEEauthorblockN{Hongliang Tian}
\IEEEauthorblockA{%
\textit{Ant Group, China}\\
tate.thl@antgroup.com}
\and
\IEEEauthorblockN{Cuiyun Gao}
\IEEEauthorblockA{%
\textit{Harbin Institute of Technology}\\
\textit{Shenzhen, China}\\
gaocuiyun@hit.edu.cn}
\thanks{\textsuperscript{*}Co-first authors.}
}

\maketitle

\begin{abstract}
Legacy C codebases remain central to system software and performance-critical infrastructure, while their weak memory-safety guarantees continue to impose substantial security and maintenance risks.
Recent C-to-Rust translation systems, including translators based on large language models (LLMs), have made automated migration increasingly feasible.
However, LLM-based translation still faces several challenges. 
Existing methods often preserve the original C skeleton, making it difficult to recover project intent and module functionalities at repository scale. 
The translated Rust code may still rely on raw pointers and unsafe constructs.
Cross-file interfaces and dependency relations are also difficult to maintain consistently across generated Rust modules. 
Even when initial translation succeeds, the resulting project still requires costly compilation repair and end-to-end validation, while naively accumulated context during this process can degrade repair stability and efficiency.

To address these challenges, we propose \texttt{SpeC2Rust}, a specification-driven pipeline for repository-level C-to-Rust translation. 
\texttt{SpeC2Rust} consists of three stages:
(1) \emph{Project-Intent Structuring}, which decomposes the C repository into semantically coherent modules and generates persistent repository-level and module-level specification documents;
(2) \emph{Spec-Guided Generation}, which translates each Rust module against these specifications with a live symbol registry to maintain cross-file consistency instead of collapsing the translated system into poorly structured monolithic Rust files;
(3) \emph{Layered-Feedback Repair}, which organizes repair into targeted batches and injects only failure-relevant diagnostics, specification context, and source evidence to support efficient compilation and test-driven correction.

We evaluate \texttt{SpeC2Rust} on eight repository-level C projects. 
On eight repository-level C projects, SpeC2Rust generates Rust projects that compile and pass the original end-to-end tests.
The generated code contains no raw-pointer declarations, raw-pointer dereferences, or unsafe lines.
On the projects shared with the compared LLM-based baseline, \texttt{SpeC2Rust} reduces cost by 66.1\% and input-token usage by 84.7\%.
\end{abstract}


\section{Introduction}\label{sec:introduction}

Modernizing legacy C code is a long-standing problem in software engineering. 
C provides weak memory-safety guarantees and a limited type system, 
which lead to high maintenance cost and frequent security vulnerabilities~\cite{SoK}. 
By contrast, Rust is increasingly regarded as a practical replacement for C because of its ownership and borrowing mechanisms, together with its strong memory-safety guarantees~\cite{RustBelt}.

Rule-based translation tools such as \texttt{C2Rust}~\cite{c2rust} can quickly translate C code into Rust. 
However, prior studies show that eliminating \texttt{unsafe} code while preserving translation quality remains difficult in practice~\cite{aliasing,towards}. 
In parallel, LLMs have demonstrated strong capability in software engineering tasks, especially code understanding, code generation~\cite{Evaluating}, and program repair~\cite{Automated}. 
Systems such as \texttt{C2SaferRust}~\cite{c2saferrust} and \texttt{SmartC2Rust}~\cite{smartc2rust} use LLMs to generate safer Rust code, but LLM-based translation still faces several major challenges.

% Place this double-column figure early so IEEEtran can float it to the top of page 2.
\begin{figure*}[!t]
  \centering
  \begin{minipage}[t]{0.26\textwidth}
    \begin{lstlisting}[language=Rust, firstnumber=1,escapeinside={(*@}{@*)}]
// Original C decomposition
void next() {}
void expr(int lev) {}
void stmt() {}
int main(int argc, char **argv) {}

// Skeleton-preserving Rust output
#[no_mangle]
pub (*@\codebad{unsafe extern "C"}@*) fn next() {}
#[no_mangle]
pub (*@\codebad{unsafe extern "C"}@*) fn expr(
    mut lev: libc::c_longlong) {}
#[no_mangle]
pub (*@\codebad{unsafe extern "C"}@*) fn stmt() {}

unsafe fn main_0(
    mut argc: libc::c_longlong,
    mut argv: (*@\codebad{*mut *mut libc::c\_char}@*),
) -> libc::c_longlong {}
pub fn main() {}
\end{lstlisting}
\refstepcounter{lstlisting}\label{lst:c-structure}
\captionof*{lstlisting}{(a) Skeleton-preserving structure}
  \end{minipage}\hfill
  \begin{minipage}[t]{0.33\textwidth}
    \begin{lstlisting}[language=Rust, firstnumber=1,escapeinside={(*@}{@*)}]
/// Lexer state remains a global raw pointer
#[no_mangle]
pub static mut (*@\codebad{p: *mut libc::c\_char}@*) =
    0 as *const libc::c_char as *mut libc::c_char;
#[no_mangle]
pub unsafe extern "C" fn next() {
    loop {
        tk = (*@\codebad{*p}@*) as libc::c_longlong;
        p = (*@\codebad{p.offset(1)}@*);
        ...
        else if is_alpha(tk) {
            (*@\codebad{*id.offset(Name)}@*) = pp as _;
        }
        else if /* Conditions */ {
            while /* Conditions */ {
                (*@\codebad{*data = *p}@*) as _;
            }
        }
        ...
}}
\end{lstlisting}
\refstepcounter{lstlisting}\label{lst:c2safer-next}
\captionof*{lstlisting}{(b) Pointer-centric implementation}
  \end{minipage}\hfill
  \begin{minipage}[t]{0.37\textwidth}
    \begin{lstlisting}[language=Rust, firstnumber=1,escapeinside={(*@}{@*)}]
#[derive(Clone,Debug,PartialEq,Eq)]
pub struct C4 {
    (*@\codegood{source: String}@*),
    (*@\codegood{position: usize}@*),
    // Other fields...
}
#[derive(Clone,Debug,PartialEq,Eq)]
pub struct C4Error {
    line: usize,
    (*@\codegood{message: String}@*),
}

impl C4 {
    pub fn next((*@\codegood{\&mut self}@*)) -> (*@\codegood{Result<(), C4Error>}@*) {}
    pub fn expr(&mut self,lev:i32)->Result<(), C4Error>{}
    pub fn stmt(&mut self)->Result<(), C4Error>{}
    fn (*@\codegood{scan\_number}@*)((*@\codegood{\&mut self}@*)) -> (*@\codegood{Result<(), C4Error>}@*) {}
    fn precedence(tok: Token)->i32{}
    // Other funcs ...
}
\end{lstlisting}
\refstepcounter{lstlisting}\label{lst:our-structure}
\captionof*{lstlisting}{(c) \texttt{SpeC2Rust}: Rust-native redesign}
  \end{minipage}
  \caption{Code comparison from a real C interpreter (\texttt{c4}) in our evaluation dataset. Light red background highlights mark C-centric structure and unsafe carryover in skeleton-preserving translation, while light teal highlights mark the Rust-native abstractions recovered by \texttt{SpeC2Rust}.}
  \label{fig:code-comparison}
\end{figure*}

Project-level C-to-Rust migration raises four challenges that are not adequately addressed by existing approaches:

\begin{itemize}
\item \textbf{C Project Intent.} LLM-generated Rust code may pass end-to-end tests while still being semantically incomplete or even degenerating into a pseudo implementation. 
In such cases, the translated result no longer preserves the original algorithmic logic of the C project.
\item \textbf{Memory Safety.} Previous approaches often construct a Rust skeleton that closely matches the original C project layout, or directly perform function-level translation over the \texttt{unsafe} Rust generated by \texttt{C2Rust}. 
As a result, the translated Rust code remains centered around \texttt{unsafe} pointers and blocks, with limited memory-safety guarantees.
\item \textbf{Cross-file Dependencies.} Rust projects are organized around modules with explicit dependency relationships. 
Existing methods sometimes place most or all translated Rust code into a single \texttt{.rs} file. 
Although this design avoids dependency management across files,
it yields poor code organization and weak maintainability. 
Moreover, as the size of the C project increases, the translation success rate tends to decrease~\cite{Scalable}.
\item \textbf{Program Repair.} The translated Rust project must be repaired through compilation and testing to ensure behavioral consistency with the original C project. Current methods often feed compiler outputs together with Rust source files back to the LLM, which causes excessive context accumulation and reduces repair effectiveness.
\end{itemize}

To address these challenges, we present \texttt{SpeC2Rust}, a specification-driven pipeline for repository-level C-to-Rust translation. 
The main idea of \texttt{SpeC2Rust} is to explicitly externalize repository-level knowledge before code generation, 
preserve it as reusable specification artifacts, and reinject it on demand during subsequent generation and repair. 

During repair, \texttt{SpeC2Rust} organizes compilation diagnostics and test failures into structured repair batches, and uses iterative repair to preserve semantic consistency between the translated Rust project and the original C project. To prevent the repair model from bypassing real failures by directly modifying the tests,  dataset test scripts are treated as read-only artifacts throughout repair. 
To address the four challenges above, \texttt{SpeC2Rust} uses three technical components. 
First, \emph{Project-Intent Structuring} partitions the repository into functional modules and generates repository-level and module-level specifications from the recovered project facts. 
Second, \emph{Spec-Guided Generation} translates each target module against these specifications and uses a live symbol registry with demand-driven evidence retrieval to maintain cross-file consistency.
Third, \emph{Layered-Feedback Repair} resolves build and test failures through targeted evidence injection, 
structured repair batching, and read-only test boundaries, thereby improving repair focus and preventing spurious success through test modification.

We evaluate \texttt{SpeC2Rust} on eight GitHub C projects. 
Across all eight projects, \texttt{SpeC2Rust} produces compilable Rust translations that pass the original end-to-end test scripts and contain no raw-pointer declarations, raw-pointer dereferences, or unsafe lines of code.
On the five projects shared with \texttt{SmartC2Rust}, \texttt{SpeC2Rust} reduces monetary cost by 66.1\% and input-token usage by 84.7\%.

Our contributions are summarized as follows:

\begin{itemize}
\item \textbf{Specification-driven Pipeline for repository-level C-to-Rust translation.} We propose \texttt{SpeC2Rust}, a three-stage workflow consisting of \emph{Project-Intent Structuring}, \emph{Spec-Guided Generation}, and \emph{Layered-Feedback Repair}, which externalizes project intent before code generation and reuses it throughout translation and repair.

\item \textbf{Specification-mediated generation and layered repair.} \texttt{SpeC2Rust} translates against reusable specification documents rather than raw source fragments alone, and repairs Rust projects through targeted knowledge reinjection with read-only test boundaries, improving safety, cross-file consistency, and repair focus.

\item \textbf{Clearer and safer Rust project reconstruction.} By breaking away from rigid C-oriented skeletons, \texttt{SpeC2Rust} produces Rust projects with clearer module organization, more idiomatic abstractions, and higher safety in the final translated repositories.

\item \textbf{Repository-level empirical validation.} On eight open-source C projects, \texttt{SpeC2Rust} produces compilable Rust translations that pass the original end-to-end test scripts while eliminating raw-pointer declarations/dereferences and unsafe lines of code. 
On the five projects shared with \texttt{SmartC2Rust}, it further reduces monetary cost by 66.1\% and input-token usage by 84.7\%.
\end{itemize}

\section{Motivating Example}

Existing skeleton-based C-to-Rust translation tools typically construct a Rust project skeleton aligned with the original C repository and then fill that skeleton with translated code.
This design preserves FFI-style function boundaries, raw-pointer-based state, and global mutable variables, thereby carrying a large amount of \texttt{unsafe} code into the generated Rust project~\cite{Laertes,CRustS}.
Moreover, because translation is performed mainly at the function and file levels, the resulting Rust code often fails to recover the module organization and interface structure required by an idiomatic Rust project~\cite{TransCoder,TypeMigratingC2Rust}.

Figure~\ref{fig:code-comparison} illustrates these limitations using code from \texttt{c4}~\cite{c4}, a small C compiler in our evaluation dataset. Listing~\ref{lst:c-structure} shows the Rust project structure produced by a skeleton-based translation strategy. 
The translated Rust layout remains closely aligned with the original C project and still contains a large amount of \texttt{unsafe} code.

Listings~\ref{lst:c2safer-next} and~\ref{lst:our-structure} further compare the internal implementation of the lexer function \texttt{next()} in \texttt{c4}. 
This function parses the input character stream into tokens consumed by the compiler. 
In Listing~\ref{lst:c2safer-next}, the translated lexer logic is still organized around the global raw pointer \texttt{p}, \texttt{extern "C"} linkage, and pervasive \texttt{unsafe} operations. 
The generated Rust therefore remains tied to C-style control flow and memory-access patterns. 
The light-red spans in the figure mark C-derived artifacts that remain in the translated Rust code.~\cite{Crown}.

By contrast, Listing~\ref{lst:our-structure} shows the Rust code generated by \texttt{SpeC2Rust}. 
The same functionality is reorganized around a \texttt{C4} struct and \texttt{C4Error} type. 
Global state is maintained through struct fields, operations are rewritten as methods over \texttt{\&mut self}, and helper functions such as \texttt{scan\_number()} improve readability and modularity. 
Failure cases are exposed through \texttt{Result}-typed interfaces rather than unstructured low-level control flow. 
This restructuring is closer to the higher-level abstraction strategies observed in human C-to-Rust migration~\cite{CToRustUserStudy}.
The light teal spans mark the Rust-native abstractions translated by \texttt{SpeC2Rust}.

This example motivates the central claim of \texttt{SpeC2Rust}: repository-level C-to-Rust translation requires more than local code conversion.  
Prior work on safer C-to-Rust migration also shows that low-level C representations often need to be lifted into Rust-native abstractions rather than translated syntactically~\cite{Crown,TaggedUnionRust}.
Such translation requires explicit recovery of project intent, specification-guided generation beyond the source skeleton, and structured repair driven by focused evidence and executable validation.

\section{Approach}\label{sec:overview}

\noindent In this section, we propose \texttt{SpeC2Rust}, a specification-driven C-to-Rust translation workflow. 
As shown in Figure~\ref{fig:pipeline-overview}, \texttt{SpeC2Rust }comprises three stages: 
(1) Project-Intent Structuring, which converts whole-project C analysis results into persistent specification documents and structured module plans, 
(2) Spec-Guided Generation, which uses specification documents to generate the Rust repository in a dependency-aware manner, and 
(3) Layered-Feedback Repair, which uses compiler diagnostics and test feedback to repair the generated project.

\begin{figure*}[!t]
\centering
\includegraphics[width=0.90\textwidth]{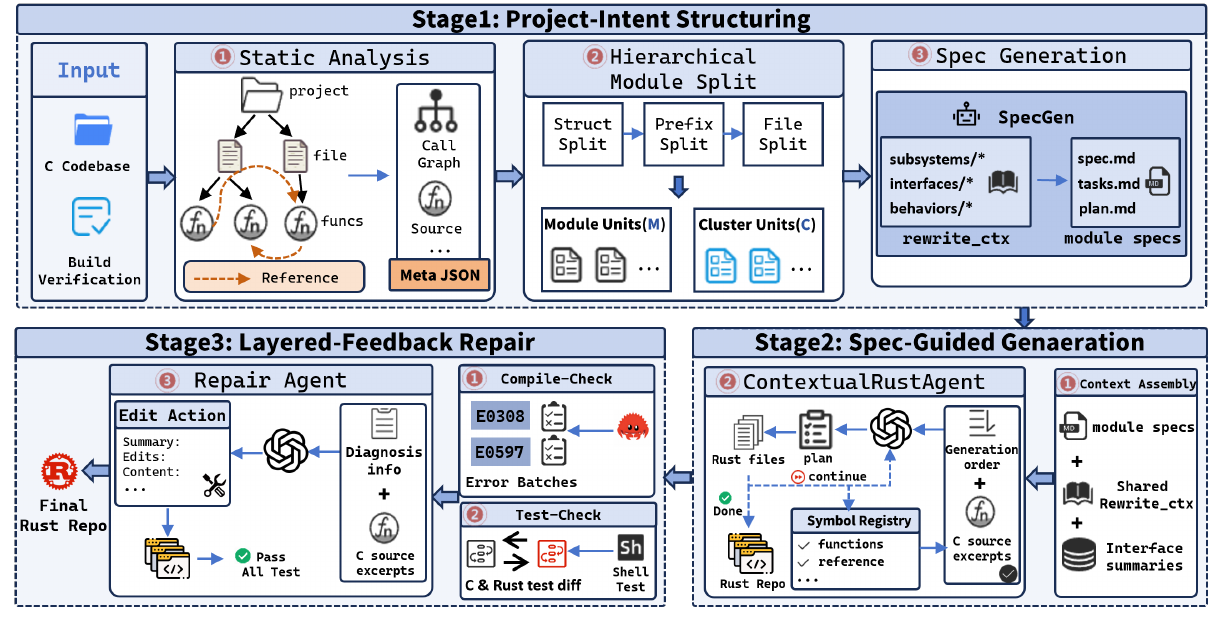}
\caption{Overview of the \texttt{SpeC2Rust} translation pipeline.}
\label{fig:pipeline-overview}
\end{figure*}

% \subsection{\textbf{Stage 1: Project-Intent Structuring}}

\subsection{Project-Intent Structuring}

% Stage 1 analyzes the input C repository and transforms repository-level knowledge embedded in the codebase into reusable specification documents. 
% This stage consists of three steps: 
% \emph{Static Analysis}, \emph{Hierarchical Module Split}, and \emph{Spec Generation}. 

Stage 1 analyzes the input C repository and produces structured specification documents that encode the project-level design intent, used by the subsequent generation and repair stages. 
It consists of three steps: \emph{Static Analysis}, \emph{Hierarchical Module Split}, and \emph{Spec Generation}.

\subsubsection{Static Analysis}
We parse each C source and header file with a tree-sitter parser, collecting per-function records (defining file, source span, body, and call sites) and per-file evidence (data definitions, global declarations, macros). We then merge per-file records into a project-level representation, resolving call sites against the function table to build a call graph and linking data-definition evidence with file ownership to form a fact graph. We also serialize all collected evidence as a JSON metadata file so that generation and repair stages can retrieve the original C context without re-parsing.

% The static analysis step takes the root directory of the C repository as input and produces two intermediate artifacts: a persistent JSON metadata file and an in-memory project fact graph used by later stages. 
% \texttt{SpeC2Rust} first enumerates all C source and header files and parses each file with a tree-sitter C parser. 
% It then traverses the syntax trees and organizes the parsed repository around function-level records. 
% For each function, it records the defining file, source span, function body, and direct call sites observed in the function body. 
% In parallel, \texttt{SpeC2Rust} records file-level evidence for data definitions, global declarations, macros, and source-file mappings.

% After all files are parsed, \texttt{SpeC2Rust} merges the per-file records into a project-level representation. 
% It resolves direct call sites against the collected function table and builds a call graph whose nodes are functions and whose edges represent observed caller-callee relations.
% The in-memory fact graph further links these function relations with file ownership and data-definition evidence, which is used by module splitting and specification generation. 
% The same source evidence is serialized as JSON metadata so that later generation and repair stages can retrieve the original C context when needed.

\subsubsection{Hierarchical Module Split}

We decompose the repository into translation-oriented module units, treating each top-level directory as an initial candidate. We then evaluate each candidate for cohesion using call graph statistics from static analysis. For a candidate module $m$ with $e_{\mathrm{in}}(m)$ intra-module and $e_{\mathrm{out}}(m)$ inter-module call edges, the cohesion score is defined as
\[
\kappa(m)=\begin{cases} 
1, & e_{\mathrm{in}}(m)=0\wedge e_{\mathrm{out}}(m)=0,\\
\dfrac{e_{\mathrm{in}}(m)}{e_{\mathrm{in}}(m)+e_{\mathrm{out}}(m)},
& \text{otherwise.}
\end{cases}
\]
Candidates whose score exceeds a threshold are materialized directly as module units. The remaining candidates are refined using three strategies applied in sequence. \textbf{Struct split} groups functions that access the same struct type. \textbf{Prefix split} groups functions sharing a common name prefix (e.g., \texttt{ast\_*}, \texttt{ir\_gen\_*}), capturing the naming conventions C projects use to delimit logical subsystems. \textbf{File-local split} groups remaining functions by source file. We chunk oversized groups further to stay within LLM context size limits. The output consists of module units for specification generation and cluster units, which are finer-grained slices of module units used as focused context slices during generation and repair. Algorithm~\ref{alg:module-splitting} summarizes the full procedure.

\begin{algorithm}[!t]
\caption{Hierarchical Module Decomposition}
\label{alg:module-splitting}
\footnotesize
\begin{algorithmic}[1]
\REQUIRE File inventory $P$, analysis metadata $A=(F,S)$, dependency graph $G=(G_{\mathrm{call}},G_{\mathrm{struct}})$
\ENSURE Module units $\mathcal{M}$ and cluster units $\mathcal{C}$
\STATE $\widehat{F}\gets \textsc{normalize\_functions}(F)$; $\widehat{S}\gets \textsc{normalize\_structs}(S)$
\STATE $\mathcal{M}^{(0)}\gets \textsc{directory\_candidates}(P)$
\STATE $\mathcal{M}\leftarrow \emptyset$
\FORALL{$m \in \mathcal{M}^{(0)}$}
    \STATE $F_m\gets \textsc{attach\_functions}(m,\widehat{F})$
    \STATE $S_m\gets \textsc{attach\_structs}(m,\widehat{S})$
    \STATE $(e_{\mathrm{in}}(m),e_{\mathrm{out}}(m),\kappa(m))\gets \textsc{compute\_cohesion}(F_m,G_{\mathrm{call}})$
    \IF{$\phi(m)=1$}
        \STATE $\mathcal{B}_1\gets \textsc{struct\_split}(F_m,G_{\mathrm{struct}})$
        \STATE $\mathcal{B}_2\gets \textsc{prefix\_split}\!\left(F_m\setminus \bigcup\mathcal{B}_1\right)$
        \STATE $\mathcal{B}_3\gets \textsc{file\_local\_split}\!\left(F_m\setminus \bigcup\mathcal{B}_1\setminus \bigcup\mathcal{B}_2\right)$
        \STATE $\mathcal{B}\gets \mathcal{B}_1\cup \mathcal{B}_2\cup \mathcal{B}_3$
        \STATE $\mathcal{B}\gets \textsc{chunk\_oversized}(\mathcal{B};\tau_g,\tau_\ell)$
        \STATE $\mathcal{M}\gets \mathcal{M}\cup \textsc{materialize\_modules}(m,\mathcal{B},G_{\mathrm{struct}})$
    \ELSE
        \STATE $\mathcal{M}\leftarrow \mathcal{M}\cup\{\textsc{attach\_module\_payload}(m,F_m,S_m)\}$
    \ENDIF
\ENDFOR
\STATE $\mathcal{M}\gets \textsc{refresh\_metrics}(\mathcal{M},G_{\mathrm{call}})$
\STATE $\mathcal{C}\gets \textsc{derive\_clusters}(\mathcal{M};\tau_g,\tau_\ell)$
\RETURN $(\mathcal{M},\mathcal{C})$
\end{algorithmic}
\end{algorithm}

% Algorithm~\ref{alg:module-splitting} illustrates this process. 
% It starts from coarse repository candidates, attaches static-analysis evidence, refines candidates when necessary, and outputs module and cluster units for specification generation.

\subsubsection{Spec Generation}

We convert each module unit into a three-document specification bundle. 
The bundle includes a requirement document covering translation obligations, an implementation plan for Rust-side organization, and a task-decomposition document listing executable translation steps. 
At the project level, it produces a rewrite context capturing repository structure, subsystem boundaries, public interfaces, and behavioral constraints. 
It also produces a translation contract defining the expected Rust-side generation boundary. 
During generation, we retrieve the relevant specification rather than passing raw C source to the model. During repair, the same documents supply source-grounded context for interpreting compiler diagnostics and test failures.
Figure~\ref{fig:spec-docs} shows example produced documents.

\begin{figure}[t]
  \centering
  % \captionsetup{skip=1pt}
  \includegraphics[width=\columnwidth]{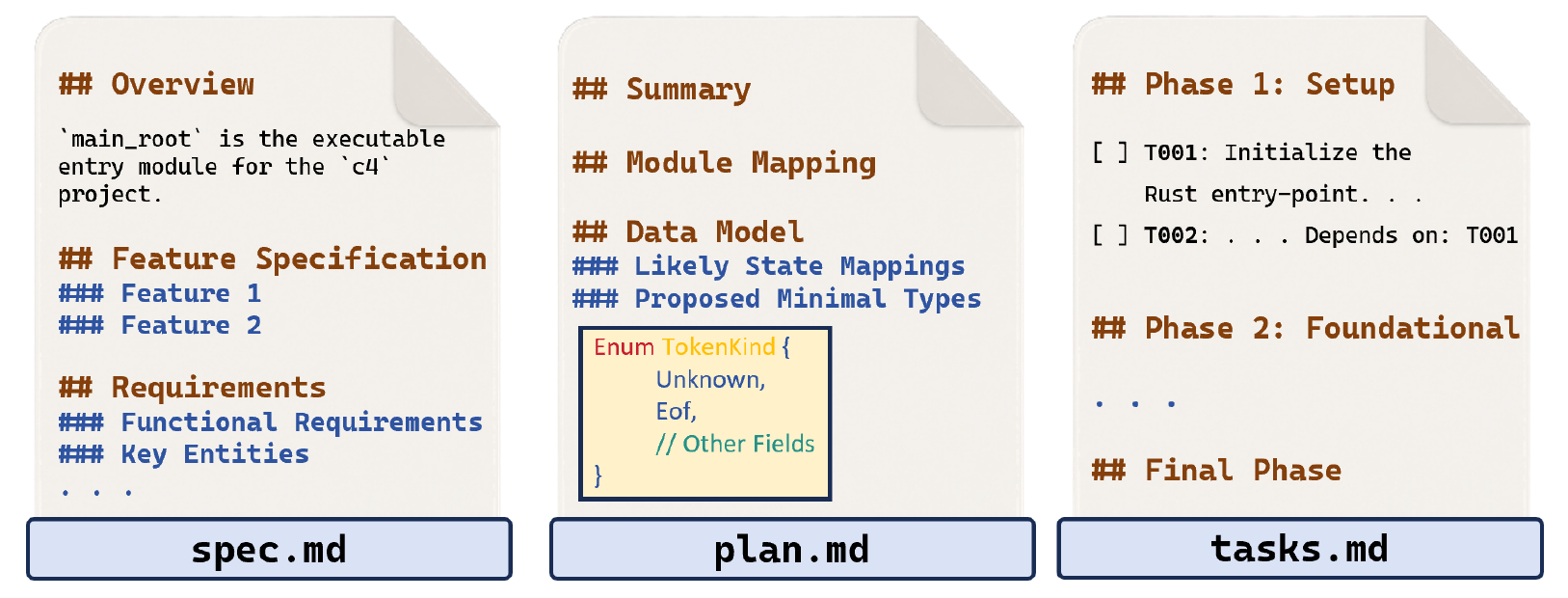}
  \caption{Example \texttt{SpeC2Rust} specification documents produced during Spec Generation.}
  \label{fig:spec-docs}
\end{figure}

\subsection{Spec-Guided Generation}

Stage 2 translates the project into Rust by using the Stage~1 specification documents as the primary context.
It consists of two steps: \emph{Context Assembly} and \emph{Rust Code Generation}.

\subsubsection{Context Assembly} 
% Before generation, we derive a file-level generation plan from the specification documents, translation contract, and source evidence. 
% The plan assigns each target Rust file a translation role, its source-based evidence, and its dependencies on other planned files. We then use the plan to retrieve a compact context package for each file, capturing module intent, interface constraints, and boundary requirements. 
% The output is a target-specific prompt context, so the generation is conditioned on the current file rather than on the entire repository.
We derive a file-level generation plan from the specification documents, translation contract, and source evidence. The plan specifies each target Rust file's generation order, the C source files it covers, and its dependencies on other planned Rust files. For each file, we assemble a focused context package combining the relevant module specification, the project-level rewrite context, and interface summaries of its dependencies.

% Before generation, \texttt{SpeC2Rust} derives a file-level generation plan from the specification documents, translation contract, and source evidence.
% The plan specifies each target file's translation role, source evidence, and dependencies on other planned files.
% \texttt{SpeC2Rust} uses this plan as the key for context selection.
% For each planned file, it selects the module specifications, interface evidence, and boundary constraints that are relevant to that file.
% The selected evidence is assembled into a target-specific context, so each prompt is conditioned on the current file rather than on the entire repository.

\subsubsection{Rust Code Generation}

We generate Rust files in the dependency order defined by the file plan, using the assembled context package as the primary input. When additional evidence is needed, \texttt{ContextualRustAgent} retrieves C source excerpts or previously generated Rust files to supplement the context. When generation reaches the model's output limit, we append the produced content as a prefix and resume generation, allowing large files to be completed across multiple passes. After generation, we extract the file's exported function signatures and referenced symbols as an interface summary and register them in a live symbol registry. Subsequent files query the registry to reuse earlier results without rereading the full Rust repository. We then run a lint checker and apply lightweight structural fixes, resolving issues such as duplicate definitions and cross-file reference errors, before advancing to the next file. Figure~\ref{fig:registry-contents} illustrates the registry format.

\begin{figure}[t]
\centering
\begin{lstlisting}[numbers=none,basicstyle=\ttfamily\scriptsize,xleftmargin=0pt,xrightmargin=0pt,framexleftmargin=0pt,framesep=4pt,linewidth=\linewidth,escapeinside={(*@}{@*)}]
(*@\textbf{Generated Rust symbol registry:}@*)
(*@\textbf{visibility policy:}@*) xxxxx

(*@\textbf{- src/c4.rs => crate::c4}@*)
  (*@\textbf{types:}@*) C4
  (*@\textbf{constants:}@*) NUM, FUN, xxx
  (*@\textbf{fields C4:}@*) src_bytes, pos, xxx
  (*@\textbf{impl C4:}@*) new, next, xxx
  (*@\textbf{references:}@*)
    - pub type C4: path=src/c4.rs
    - field C4::src_bytes: Vec<u8>: path=src/c4.rs
    - const NUM: path=src/c4.rs
    - pub fn C4::main(...) -> i32: path=src/c4.rs

(*@\textbf{- src/main.rs}@*)
  (*@\textbf{functions:}@*) main
  (*@\textbf{references:}@*)
    - fn main(): path=src/main.rs
\end{lstlisting}
\caption{An example of the generated Rust symbol registry used in Stage~2.}
\label{fig:registry-contents}
\end{figure}

% When the selected specification context is insufficient, \texttt{ContextualRustAgent} requests additional evidence on demand. 
% The requested evidence may come from specification documents, original C source records, previously generated Rust files, the symbol registry, or the file plan. 
% After each file is generated, \texttt{SpeC2Rust} checks it against the translation boundary and updates the registry before generating the next file. 
% Figure~\ref{fig:registry-contents} illustrates the registry format.

\subsection{Layered-Feedback Repair}

Stage~3 repairs the generated Rust project using compiler diagnostics and
C/Rust behavioral differences, drawing on Stage~1 specifications and Stage~2 interface summaries as repair context. It consists of four steps:
\emph{execution environment alignment}, \emph{structural normalization},
\emph{compilation-Guided Repair}, and \emph{test-driven repair closure}.
% Stage~3 repairs the Rust project generated in Stage~2 using compiler diagnostics and C/Rust behavioral differences. 
% It reuses Stage~1 specifications and Stage~2 interface summaries as repair context. 
% This stage consists of four steps: \emph{Execution Environment Alignment}, \emph{Structural Normalization}, \emph{Compilation-Guided Repair}, and \emph{Test-Driven Repair Closure}. 

% These steps form an iterative loop.
% In each iteration, we collect feedback from aligned execution, localize affected files, construct layered repair prompts, apply generated patches, and rerun validation. 
% The output is a Rust project that compiles and passes the original end-to-end tests.

% Stage 3 repairs the Rust program generated in Stage 2 through compiler and test feedback. 
% Given a translated Rust program, the original C program, a test script, and project knowledge, the system performs repair under an aligned execution environment. 
% The repair process uses compiler diagnostics and behavioral differences as feedback signals.

% This stage consists of four components: \emph{execution environment alignment}, \emph{structural normalization}, \emph{compilation-guided repair}, and \emph{test-driven repair closure}. 
% These components form an iterative repair loop. 
% In each iteration, \texttt{SpeC2Rust} collects feedback, localizes the affected files, builds a layered repair prompt, and invokes the repair agent to generate patches.

\subsubsection{Execution Environment Alignment}

We construct an aligned execution environment for the C project and the Rust project, recording dependency settings, working directories, input/output conventions, and test commands needed to run both under the same setup. This environment provides the basis for collecting compiler diagnostics and C/Rust behavioral differences used in the subsequent repair steps.

% We construct an aligned execution environment for the original C project and the translated Rust project. 
% The environment records execution conventions (dependency settings, working directories, input/output rules, and test commands) needed to run both projects under the same setup. 
% We then collect execution-grounded feedback from this environment, including compiler diagnostics and C/Rust behavioral differences. 
% We use the resulting feedback records as repair inputs by the following steps.

% \textbf{Execution Environment Alignment.} Stage 3 builds an execution environment for the translated Rust project, which is aligned with the original C project. It records dependency settings, working directories, input and output rules, and test execution commands.

% Both the C program and the Rust program run under the same test setup. 
% Compilation diagnostics, runtime errors, and behavior differences are collected from this aligned environment. 
% These results are used as repair feedback in later components.

\subsubsection{Structural Normalization}

Before model-based repair, we apply rule-based preprocessing to the generated Rust project to remove structural artifacts that can be resolved without model inference, including formatting errors, invalid module declarations, residual syntax fragments, and import issues. This allows later repair rounds to focus on substantive errors such as type mismatches, ownership violations, and interface mismatches.

% Before model-based repair, we apply lightweight rule-based preprocessing to the generated Rust project. 
% We removes rule-recoverable structural artifacts, such as formatting errors, invalid module declarations, residual syntax fragments, and import issues. 
% The output is a normalized Rust project, allowing later repair rounds to focus on type errors, ownership constraints, cross-module dependencies, and interface adaptation.

% \textbf{Structural Normalization.} Structural normalization takes the generated Rust code as input. 
% It applies lightweight rule-based preprocessing before model-based repair. 
% The preprocessing handles formatting errors, incorrect module declarations, remaining syntax fragments, and import issues that can be fixed by rules.

% The output is a normalized Rust project in which rule-recoverable structural problems have been removed before model-based repair. 
% The following repair loop mainly handles type checking, ownership constraints, cross-module dependencies, and interface adaptation.

\subsubsection{Compilation-Guided Repair}

We run \texttt{cargo check} and \texttt{cargo build} to collect compiler diagnostics and group related diagnostics into repair batches by affected files and dependency relations. For each batch, we build a diagnosis plan identifying the Rust units, module interfaces, and configuration files involved in the failure.

% We run \texttt{cargo check} and \texttt{cargo build} to collect compiler diagnostics, then group related diagnostics into repair batches according to affected files and dependency relations. 
% For each active batch, we build a diagnosis plan that identifies the Rust units, module interfaces, dependencies, and configuration files involved in the failure. 
% The repair loop is organized into major and minor rounds. Each minor round repairs active batches, while each major round maintains repair memory and controls rollback when progress stalls.

We organize the repair loop into major and minor rounds. Minor rounds iterate over active batches. Major rounds maintain repair memory and trigger rollback when progress stalls. \texttt{RustRepairAgent} repairs each batch using a layered prompt with five components. The repair protocol layer encodes stable repair rules. The failure instance layer describes the current batch using minimized diagnostics and candidate causes. The project knowledge layer provides specifications, module responsibilities, and inter-file dependencies from earlier stages. The evidence retrieval layer supplies Rust code slices, C source excerpts, and execution traces for the current diagnosis. The repair memory layer records the current analysis, attempted repair directions, and edit summaries across rounds. Figure~\ref{fig:stage3-repair-prompt} shows this prompt organization. After each patch, we rerun compilation and update the repair memory with the remaining diagnostics and applied edits. If a major round reaches its repair limit without sufficient progress, we restore the code snapshot from the start of that round and retry.

\begin{figure}[t]
\centering
\begin{lstlisting}[numbers=none,basicstyle=\ttfamily\scriptsize,xleftmargin=0pt,xrightmargin=0pt,framexleftmargin=0pt,framesep=4pt,linewidth=\linewidth,escapeinside={(*@}{@*)}]
(*@\textbf{Layered repair prompt (Stage 3)}@*)

(*@\textbf{[Invariant repair protocol]}@*)
  translation objective and repair role
  read-only test / fixture conventions
  allowed requests, edit scope, output schema

(*@\textbf{[Failure instance]}@*)
  active batch or failing script summary
  minimized compiler diagnostics / test diffs
  candidate features, files, execution clues

(*@\textbf{[Persistent project knowledge]}@*)
  specification documents from Stages 1--2
  module responsibilities / symbol index
  reusable project structure and intent

(*@\textbf{[On-demand evidence injection]}@*)
  Rust line slices for target files
  C function excerpts or file ranges
  test artifacts, rerun traces, raw diffs

(*@\textbf{[Repair memory and validation]}@*)
  current-round summary
  cross-round handoff summary
  structured edits + follow-up reads
  cargo / test rerun + refreshed summary
\end{lstlisting}
\caption{Prompt organization in Stage~3 (\emph{Layered Equivalence Closure}). A concrete
script-level repair case is abstracted into layered prompt components: invariant repair protocol, failure-instance context, persistent project knowledge, on-demand evidence, and compact repair memory. This layered knowledge injection keeps the active context focused on the current failure while avoiding uncontrolled prompt growth.}
\label{fig:stage3-repair-prompt}
\end{figure}

\subsubsection{Test-Driven Repair Closure}

After the Rust project compiles, we run the original test scripts on both the C reference and the Rust program. We compare their execution results, including outputs, exit status, runtime failures, and assertion failures, and convert each behavioral difference into a failure instance. We then construct a compact repair input from minimized test differences, execution traces, and relevant C/Rust code evidence. The repair agent combines this feedback with project knowledge and source evidence to fix behavioral mismatches such as boundary conditions, integer behavior, error paths, and cross-module calls. The loop terminates when the Rust project compiles and passes all end-to-end tests.

\section{Experimental Setup And Implementation}\label{sec:exp-setup}

\subsubsection{Dataset.}
We use C projects from prior work on project-scale C-to-Rust translation, primarily following the evaluation datasets used by prior work~\cite{smartc2rust,c2saferrust}. 
Our dataset contains eight projects: \texttt{c4}~\cite{c4}, \texttt{yank}~\cite{yank}, \texttt{sds}~\cite{sds}, \texttt{which}~\cite{which}, \texttt{cat}~\cite{cat},
\texttt{pwd}~\cite{pwd}, \texttt{shc}~\cite{shc}, and \texttt{cflow}~\cite{cflow}.
Their sizes range from 418 to 49,757 lines of C code. Each project is accompanied by end-to-end shell test scripts. 
Table~\ref{tab:dataset} summarizes the dataset composition. \textbf{LoC}, \textbf{Files}, and \textbf{Functions} characterize repository scale and cross-file complexity, which directly affect translation difficulty. 
\textbf{\#Test Source} indicates the provenance of the executable end-to-end test script, clarifying whether the validation artifacts come from prior translation work or from the original project itself. 
\textbf{\#Test Cases} reflects the amount of observable semantic behavior covered by each script, which is important for interpreting the strength of semantic-equivalence evaluation. 
\textbf{\#Modules} reports the number of translation modules produced by Stage~1 in \texttt{SpeC2Rust}; this value serves as an indicator of repository decomposition difficulty and shows how much internal structure must be reconstructed before generation and repair.

\begin{table}[t]
\caption{Evaluation dataset used in the experimental evaluation.}
\label{tab:dataset}
\centering
\scriptsize
\setlength{\tabcolsep}{4pt}
\renewcommand{\arraystretch}{1.10}
\resizebox{\columnwidth}{!}{%
\begin{tabular}{lrrrlrr}
\toprule
\textbf{Project} & \textbf{LoC} & \textbf{Files} & \textbf{Functions} & \textbf{\#Test Source} & \textbf{\#Test Cases} & \textbf{\#Modules} \\
\midrule
\texttt{c4}    &    986 &   2 &   4 & \texttt{SmartC2Rust} &  2 &   2 \\
\texttt{yank}  &    418 &   1 &  13 & \texttt{SmartC2Rust} &  8 &   1 \\
\texttt{sds}   &  1,110 &   4 &  47 & \texttt{SmartC2Rust} & 24 &   1 \\
\texttt{which} &  2,327 &  15 &  24 & \texttt{SmartC2Rust} & 40 &   2 \\
\texttt{cat}   & 28,353 & 123 & 134 & \texttt{C2SaferRust} &  4 &  39 \\
\texttt{pwd}   & 25,790 & 101 & 129 & \texttt{C2SaferRust} &  2 &  29 \\
\texttt{shc}   &  1,253 &   1 &  18 & Source project        & 45 &   1 \\
\texttt{cflow} & 49,757 & 146 & 487 & \texttt{SpeC2Rust}   & 15 & 120 \\
\bottomrule
\end{tabular}
}
\end{table}

\subsubsection{Baselines.}
We compare \texttt{SpeC2Rust} against four representative baselines:
\begin{itemize}
\item \texttt{C2Rust}~\cite{c2rust} is a rule-based open-source toolchain for translating C code (primarily targeting the C99 standard) into Rust.
\item \texttt{C2SaferRust}~\cite{c2saferrust} takes the \texttt{unsafe} Rust code produced by \texttt{C2Rust}, first partitions the code into chunks, and then combines rule-based transformations with LLM-based rewriting to translate \texttt{unsafe} Rust into safer Rust.
\item \texttt{EvoC2Rust}~\cite{evoc2rust} is a skeleton-driven translation tool that first constructs a corresponding Rust skeleton from the input C project and then uses LLM to generate code that fills the skeleton.
\item \texttt{SmartC2Rust}~\cite{smartc2rust} is an iterative, feedback-driven C-to-Rust translation approach via large language models for safety and equivalence, and it serves as our primary comparison point.
\end{itemize}

\subsubsection{Metrics}
Following prior work on C-to-Rust translation~\cite{smartc2rust,c2saferrust,C2RustXW}, we evaluate \texttt{SpeC2Rust} from three perspectives: safety, semantic equivalence, and cost.
The same metric abbreviations are used throughout the later RQs.

\textbf{Safety metrics.} We use three safety-related metrics: unsafe lines of code (\texttt{uLoC}), raw-pointer declarations (\texttt{Decl.}), and raw-pointer dereferences (\texttt{Deref.}). 
These metrics capture how much low-level unsafe behavior remains in the translated Rust project.

\textbf{Semantic-equivalence metrics.} We report compile success (\texttt{Comp.}) and test pass status/rate (\texttt{Tests Passed}). 
Following prior work, semantic equivalence is assessed by whether the translated Rust project compiles successfully and preserves the observable behavior of the original C project under the original end-to-end test scripts.

\textbf{Cost metrics.} We report the number of model interactions (\texttt{Chats}), input tokens (\texttt{Input Tokens}), output tokens (\texttt{Output Tokens}), end-to-end latency (\texttt{Time}), and total API cost (\texttt{Cost}). 
These metrics are used to evaluate the practical overhead of repository-level translation and iterative repair.

\subsubsection{Implementation Details.}
In all LLM-based experiments, we use the \texttt{gpt-5.4} API model with \texttt{max\_tokens} set to 8192. 
For evaluation consistency, we place all test scripts under \texttt{test/} at the root of each C project and copy them into the translated Rust repository after generation. 
We keep these scripts unchanged and only redirect the executable path from the original C binary to the translated Rust binary. 
This setup lets us evaluate semantic preservation with the original project tests while avoiding any LLM-driven rewriting of the evaluation scripts.

\section{Experimental Evaluation}\label{sec:experiments}

We evaluate \texttt{SpeC2Rust} on eight C projects and compare it against representative C-to-Rust translation baselines.

\subsection{Research Questions}\label{sec:research-questions}

This section focuses on the following research questions:

\noindent\textbf{RQ1:} How effective is \texttt{SpeC2Rust} at improving translation safety compared with representative C-to-Rust translation baselines?

\noindent\textbf{RQ2:} To what extent can \texttt{SpeC2Rust} preserve repository-level semantic equivalence with the original C projects?

\noindent\textbf{RQ3:} How cost-efficient is \texttt{SpeC2Rust} compared with prior repository-level C-to-Rust translation approaches?

\noindent\textbf{RQ4:} What is the contribution of the major components of \texttt{SpeC2Rust} to the final translation performance?

\subsection{RQ1: Safety}\label{sec:rq1-safety}

Table~\ref{tab:rq1-safety} reports the safety results of \texttt{SpeC2Rust} and representative C-to-Rust translation baselines. 
Across the eight evaluated repositories, \texttt{SpeC2Rust} produces \texttt{0} raw-pointer declarations, \texttt{0} raw-pointer dereferences, and \texttt{0\%/0} unsafe code on every project; 
the aggregate row further shows a \texttt{100.0\%} reduction in all three safety metrics relative to \texttt{C2Rust}.

\begin{table*}[t]
\caption{RQ1 safety comparison. Lower values are better for all safety metrics. Boldface marks the best value in a row, underlining marks tied best values, and ``$\sim$'' indicates that the tool did not produce a successful translation or did not report the corresponding metric.}
\label{tab:rq1-safety}
\centering
\scriptsize
\setlength{\tabcolsep}{3pt}
\renewcommand{\arraystretch}{1.1}
\begin{threeparttable}
\resizebox{\textwidth}{!}{%
\begin{tabular}{ll ccc ccc ccc c ccc}
\toprule
\multirow{2}{*}{\textbf{Project}} & \multirow{2}{*}{\textbf{LoC}}
& \multicolumn{3}{c}{\textbf{C2Rust}}
& \multicolumn{3}{c}{\textbf{C2SaferRust}}
& \multicolumn{3}{c}{\textbf{SmartC2Rust}}
& \multicolumn{1}{c}{\textbf{EvoC2Rust}}
& \multicolumn{3}{c}{\textbf{SpeC2Rust}} \\
\cmidrule(lr){3-5} \cmidrule(lr){6-8} \cmidrule(lr){9-11} \cmidrule(lr){12-12}
\cmidrule(lr){13-15}
& & \textbf{Decl.} & \textbf{Deref.} & \textbf{Unsafe/uLoC}
& \textbf{Decl.} & \textbf{Deref.} & \textbf{Unsafe/uLoC}
& \textbf{Decl.} & \textbf{Deref.} & \textbf{Unsafe/uLoC}
& \textbf{Comp.}
& \textbf{Decl.} & \textbf{Deref.} & \textbf{Unsafe/uLoC} \\
\midrule
\texttt{c4}    &    502
& 201 & 309 & 89.75\%/1,883
& 91 & 307 & 89.78\%/665
& \tied{0} & \tied{0} & \tied{0\%/0}
& $\sim$
& \tied{0} & \tied{0} & \tied{0\%/0} \\
\texttt{yank}  &    465
& 261 & 23 & 71.13\%/1,081
& 148 & 10 & 40.06\%/215
& \tied{0} & \tied{0} & 0.04\%/1
& $\sim$
& \tied{0} & \tied{0} & \best{0\%/0} \\
\texttt{sds}   &  1,943
& 672 & 188 & 91.69\%/2,481
& 328 & 106 & 9.33\%/212
& \tied{0} & \tied{0} & \tied{0\%/0}
& $\sim$
& \tied{0} & \tied{0} & \tied{0\%/0} \\
\texttt{which} &  3,530
& 428 & 138 & 72.35\%/1,982
& 230 & 48 & 16.86\%/327
& \tied{0} & \tied{0} & \tied{0\%/0}
& $\sim$
& \tied{0} & \tied{0} & \tied{0\%/0} \\
\texttt{cat}   & 28,313
& 1,079 & 311 & 75.40\%/7,460
& 719 & 185 & 10.72\%/674
& $\sim$ & $\sim$ & $\sim$
& $\sim$
& \best{0} & \best{0} & \best{0\%/0} \\
\texttt{pwd}   & 25,790
& 1,019 & 286 & 71.70\%/5,859
& 650 & 147 & 10.81\%/536
& \tied{0} & \tied{0} & \tied{0\%/0}
& $\sim$
& \tied{0} & \tied{0} & \tied{0\%/0} \\
\texttt{shc}   &  1,364
& 1,994 & 41 & 51.29\%/2,940
& 1,781 & 5 & 9.88\%/293
& $\sim$ & $\sim$ & $\sim$
& $\sim$
& \best{0} & \best{0} & \best{0\%/0} \\
\texttt{cflow} & 49,757
& 6,168 & 3,315 & 82.56\%/35,413
& 4,994 & 2,590 & 14.36\%/3,575
& $\sim$ & $\sim$ & $\sim$
& $\sim$
& \best{0} & \best{0} & \best{0\%/0} \\
\midrule
\textbf{Rel. vs C2Rust} & --
& -- & -- & --
& $\downarrow$24.4\% & $\downarrow$26.3\% & $\downarrow$89.0\%
& N/A & N/A & N/A
& N/A
& \textbf{$\downarrow$100.0\%} & \textbf{$\downarrow$100.0\%} & \textbf{$\downarrow$100.0\%} \\
\bottomrule
\end{tabular}%
}
\end{threeparttable}
\end{table*}

\textbf{Raw-pointer elimination.} 
Across all eight projects, \texttt{SpeC2Rust} produces zero raw-pointer declarations and zero raw-pointer dereferences. 
In Table~\ref{tab:rq1-safety}, \texttt{C2Rust} retains between 201 and 6,168 raw-pointer declarations and between 23 and 3,315 raw-pointer dereferences per project, while \texttt{C2SaferRust} still leaves between 91 and 4,994 declarations and between 5 and 2,590 dereferences. 
On the five projects where \texttt{SmartC2Rust} reports safety outputs, it also reduces raw-pointer declarations and dereferences to zero, but it does not report completed results for \texttt{cat}, \texttt{shc}, or \texttt{cflow}. 
These results indicate that directly preserving C-oriented structure tends to carry pointer-based interfaces and shared-state patterns into the Rust output, whereas \texttt{SpeC2Rust} systematically removes them at repository scale.

\textbf{Unsafe-code removal.} 
\texttt{SpeC2Rust} achieves \texttt{0\%/0} unsafe code on all eight projects. 
In contrast, \texttt{C2Rust} ranges from \texttt{51.29\%/2,940} on \texttt{shc} to \texttt{91.69\%/2,481} on \texttt{sds}, and \texttt{C2SaferRust} still ranges from \texttt{9.33\%/212} on \texttt{sds} to \texttt{89.78\%/665} on \texttt{c4}. 
For \texttt{SmartC2Rust}, four projects finish at \texttt{0\%/0}, while \texttt{yank} still retains \texttt{0.04\%/1}. 
The aggregate row further shows that, relative to \texttt{C2Rust}, \texttt{SpeC2Rust} reduces declarations, dereferences, and unsafe lines by \texttt{100.0\%}.
This result shows that specification-driven repository reconstruction is more effective than skeleton-preserving translation at removing residual \texttt{unsafe} code.

\rqanswer{RQ1}{Compared with \texttt{C2Rust}, \texttt{SpeC2Rust} reduces raw-pointer declarations, raw-pointer dereferences, and unsafe lines of code by 100.0\% in the aggregate, while maintaining full translation coverage across the evaluation dataset.}

\subsection{RQ2: Semantic Equivalence}\label{sec:rq2-semantic}

Table~\ref{tab:rq2-semantic} summarizes the semantic-equivalence results of \texttt{SpeC2Rust}. 
Across all eight projects in the evaluation dataset, the translated Rust projects compile successfully and pass the original end-to-end test scripts. 
In contrast, \texttt{SmartC2Rust} fails to compile on four projects in the current result sheet, and \texttt{EvoC2Rust} does not produce a successful translation for any project in our dataset. 
\texttt{C2SaferRust} compiles on all projects, because each rewritten function is checked by compilation and testing before replacing its original \texttt{unsafe} counterpart; when the rewritten version fails, the pipeline falls back to the corresponding \texttt{C2Rust} output to preserve buildability. 
From the perspective of completed repository-level translations, \texttt{SpeC2Rust} is also the only method in the current evaluation that simultaneously provides full eight-project coverage and zero-unsafe outputs. 
\texttt{SmartC2Rust} reaches \texttt{0\%/0} unsafe code on four projects and \texttt{0.04\%/1} on \texttt{yank}, but it does not report completed safety results for \texttt{cat}, \texttt{shc}, or \texttt{cflow}. \texttt{EvoC2Rust} does not produce a compilable translation on any of the eight projects. 
This coverage difference is important because safety improvements are only meaningful when they are achieved on completed repository translations.

\begin{table}[t]
\caption{RQ2 semantic-equivalence results of \texttt{SpeC2Rust}. Repair Iter. denotes the number of compile-and-test repair iterations after initial Rust generation. Clippy W/E denote the numbers of warnings and errors reported by the final \texttt{cargo clippy} run.}
\label{tab:rq2-semantic}
\centering
\scriptsize
\setlength{\tabcolsep}{4pt}
\renewcommand{\arraystretch}{1.10}
\begin{threeparttable}
\resizebox{\columnwidth}{!}{%
\begin{tabular}{lrrrrc}
\toprule
\multirow{2}{*}{\textbf{Project}} & \multirow{2}{*}{\textbf{Repair Iter.}} & \multicolumn{2}{c}{\textbf{Clippy}} & \multirow{2}{*}{\textbf{Rust LoC}} & \multirow{2}{*}{\textbf{Tests Passed}} \\
\cmidrule(lr){3-4}
& & \textbf{Warning} & \textbf{Error} & & \\
\midrule
\texttt{c4}    &   4 &  14 & 0 &    992 & $\checkmark$ \\
\texttt{yank}  &  12 &  10 & 0 &    514 & $\checkmark$ \\
\texttt{sds}   &   6 &  10 & 0 &    740 & $\checkmark$ \\
\texttt{which} &   7 &  11 & 1 &  1,189 & $\checkmark$ \\
\texttt{cat}   &  22 &  93 & 1 &  3,538 & $\checkmark$ \\
\texttt{pwd}   &   7 &  30 & 0 &  2,134 & $\checkmark$ \\
\texttt{shc}   &   9 &  10 & 0 &    895 & $\checkmark$ \\
\texttt{cflow} &  65 & 114 & 0 & 11,586 & $\checkmark$ \\
\bottomrule
\end{tabular}
}
\end{threeparttable}
\end{table}

The repair effort varies substantially with project scale. 
The smallest completed case, \texttt{c4}, requires only four repair iterations, whereas \texttt{cflow} requires 65 iterations.
\texttt{cat} is the second most demanding case with 22 iterations. 
This trend is consistent with the larger code volume and richer cross-file dependencies of these repositories, which create more opportunities for build-level mismatches and test-observable semantic deviations during translation.

The final translated code also remains analyzable by standard Rust tooling. 
Most projects finish with zero \texttt{cargo clippy} errors, while \texttt{which} and \texttt{cat} each retain a single \texttt{clippy} error. 
The remaining warnings are mainly associated with code paths, helper functions, or interface fragments that are weakly exercised by the available end-to-end tests.
Nevertheless, all eight translated projects pass the original end-to-end tests, which is the primary criterion for semantic equivalence in our evaluation.

\rqanswer{RQ2}{\texttt{SpeC2Rust} achieves complete end-to-end semantic equivalence on all eight projects in the current dataset, while remaining executable and largely well-formed under the Rust toolchain.}

\subsection{RQ3: Cost}\label{sec:rq3-cost}

We evaluate the overhead of \texttt{SpeC2Rust} from two aspects: (1) performance overhead, including latency, chat count, and token usage, and (2) monetary cost in US dollars. 
Table~\ref{tab:rq3-cost} reports the results for each project. 
Since \texttt{SmartC2Rust} does not complete \texttt{cat}, \texttt{shc}, or \texttt{cflow} in our evaluation, we use the five projects completed by both tools, i.e., \texttt{c4}, \texttt{yank}, \texttt{sds}, \texttt{which}, and \texttt{pwd}, for direct comparison. 
The full eight-project result of \texttt{SpeC2Rust} is also reported to show its total overhead on the complete benchmark.

\begin{table*}[t]
\caption{RQ3 cost comparison between \texttt{SpeC2Rust} and \texttt{SmartC2Rust}. A dash indicates that \texttt{SmartC2Rust} did not complete the corresponding project successfully, so no directly comparable cost trace is available.}
\label{tab:rq3-cost}
\centering
\scriptsize
\setlength{\tabcolsep}{2.5pt}
\renewcommand{\arraystretch}{1.12}
\begin{threeparttable}
\begin{tabular}{lrrrrrrrrrr}
\toprule
\textbf{Project}
& \multicolumn{2}{c}{\textbf{Cost/\$}}
& \multicolumn{2}{c}{\textbf{Time/s}}
& \multicolumn{2}{c}{\textbf{Chats}}
& \multicolumn{2}{c}{\textbf{Input Tokens/k}}
& \multicolumn{2}{c}{\textbf{Output Tokens/k}} \\
\cmidrule(lr){2-3}
\cmidrule(lr){4-5}
\cmidrule(lr){6-7}
\cmidrule(lr){8-9}
\cmidrule(lr){10-11}
& \textbf{\texttt{SpeC}} & \textbf{\texttt{Smart}}
& \textbf{\texttt{SpeC}} & \textbf{\texttt{Smart}}
& \textbf{\texttt{SpeC}} & \textbf{\texttt{Smart}}
& \textbf{\texttt{SpeC}} & \textbf{\texttt{Smart}}
& \textbf{\texttt{SpeC}} & \textbf{\texttt{Smart}} \\
\midrule
\texttt{c4}
& \num{1.13} & \num{4.41}
& \num{548} & \num{2838}
& \num{21} & \num{28}
& \num{120.02} & \num{1092.29}
& \num{32.96} & \num{35.99} \\

\texttt{yank}
& \num{2.05} & \num{6.32}
& \num{582} & \num{2632}
& \num{25} & \num{38}
& \num{343.43} & \num{1449.77}
& \num{24.99} & \num{63.62} \\

\texttt{sds}
& \num{1.39} & \num{7.79}
& \num{768} & \num{3158}
& \num{18} & \num{37}
& \num{158.99} & \num{2365.60}
& \num{38.00} & \num{50.46} \\

\texttt{which}
& \num{2.30} & \num{11.39}
& \num{905} & \num{5058}
& \num{33} & \num{57}
& \num{357.53} & \num{3864.39}
& \num{51.76} & \num{119.48} \\

\texttt{pwd}
& \num{6.22} & \num{8.70}
& \num{3105} & \num{12352}
& \num{157} & \num{109}
& \num{817.92} & \num{2951.34}
& \num{169.57} & \num{132.68} \\

\texttt{cat}
& \num{11.32} & $-$
& \num{4939} & $-$
& \num{222} & $-$
& \num{188.76} & $-$
& \num{232.80} & $-$ \\

\texttt{shc}
& \num{2.09} & $-$
& \num{789} & $-$
& \num{22} & $-$
& \num{586.28} & $-$
& \num{52.93} & $-$ \\

\texttt{cflow}
& \num{36.50} & $-$
& \num{10626} & $-$
& \num{565} & $-$
& \num{6220.41} & $-$
& \num{738.28} & $-$ \\

\midrule
\textbf{SpeC2Rust total (8 proj.)}
& \textbf{\num{63.00}} & $-$
& \textbf{\num{22262}} & $-$
& \textbf{\num{1063}} & $-$
& \textbf{\num{8793.34}} & $-$
& \textbf{\num{1341.29}} & $-$ \\

\textbf{Overlap total (5 proj.)}
& \textbf{\num{13.09}} & \num{38.61}
& \textbf{\num{5908}} & \num{26038}
& \textbf{\num{254}} & \num{269}
& \textbf{\num{1797.89}} & \num{11723.39}
& \textbf{\num{317.28}} & \num{402.23} \\
\bottomrule
\end{tabular}
\end{threeparttable}
\end{table*}

\textbf{Performance.}
On the five common projects, \texttt{SpeC2Rust} requires 254 chats, 1,797.89\,K input tokens, 317.28\,K output tokens, and 5,908\,s in total. In comparison, \texttt{SmartC2Rust} requires 269 chats, 11,723.39\,K input tokens, 402.23\,K output tokens, and 26,038\,s. 
Thus, \texttt{SpeC2Rust} reduces latency by 77.3\%, input-token usage by 84.7\%, and output-token usage by 21.1\%. 
The chat count decreases by 5.6\%. 
The reduction is mainly reflected in input tokens and latency, rather than in the number of interactions.

\begin{figure}[t]
\centering
\begin{tikzpicture}
\begin{axis}[
  width=0.92\columnwidth,
  height=0.58\columnwidth,
  xmin=1, xmax=6,
  xtick={1,2,3,4,5,6},
  xlabel={Repair round},
  ylabel={Input tokens (K)},
  ymin=0,
  grid=both,
  grid style={line width=0.2pt, draw=gray!30},
  major grid style={line width=0.3pt, draw=gray!40},
  tick label style={font=\scriptsize},
  label style={font=\scriptsize},
  legend style={
    at={(0.5,-0.30)},
    anchor=north,
    legend columns=2,
    draw=blue!20!black,
    rounded corners=2pt,
    fill=blue!6,
    font=\scriptsize,
    inner xsep=6pt,
    inner ysep=4pt,
  },
  every axis plot/.append style={line width=1pt, mark size=2.2pt},
]
\addplot[color=blue!70!black, mark=*] coordinates {
  (1,26.44) (2,28.17) (3,25.99) (4,29.01) (5,29.02) (6,29.94)
};
\addplot[color=red!75!black, mark=square*] coordinates {
  (1,14.48) (2,41.71) (3,47.93) (4,55.74) (5,73.78) (6,75.15)
};
\legend{SpeC2Rust, SmartC2Rust}
\end{axis}
\end{tikzpicture}
\caption{Input-token consumption across repair rounds on \texttt{c4}. The legend is placed below the plot for readability.}
\label{fig:c4-repair-token}
\end{figure}
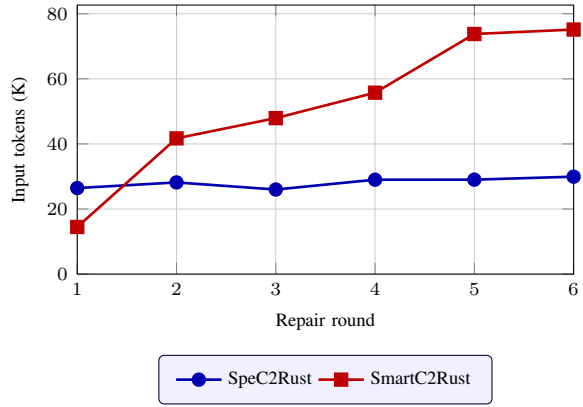

Figure~\ref{fig:c4-repair-token} further shows the input-token usage during repair on \texttt{c4}. 
Both tools require six repair rounds on this trace. 
However, after the initial round, \texttt{SpeC2Rust} keeps the per-round input size lower than \texttt{SmartC2Rust}. 
The cumulative input-token usage is 168.58\,K for \texttt{SpeC2Rust}, compared with 308.79\,K for \texttt{SmartC2Rust}, a 45.4\% reduction. 
This is because \texttt{SpeC2Rust} constructs each repair prompt from the repair protocol, the relevant specification context, and the observed failure evidence. 
It does not carry forward a growing prompt state across repair rounds.

\textbf{Cost.}
On the five common programs, \texttt{SpeC2Rust} costs \$13.09 in total, while \texttt{SmartC2Rust} costs \$38.61. 
This corresponds to a 66.1\% reduction. 
Since the pricing of LLM calls is mainly determined by input and output tokens, the cost reduction is consistent with the token reduction above. 
The difference is especially large for input tokens, because \texttt{SmartC2Rust} needs to repeatedly provide large file-level contexts during interactive repair, whereas \texttt{SpeC2Rust} uses specification-guided repair contexts.

Across all eight programs, \texttt{SpeC2Rust} completes every translation. 
Its total cost is \$63.00, with 22,262\,s of latency, 1,063 chats, 8,793.34\,K input tokens, and 1,341.29,K output tokens. 
In contrast, \texttt{SmartC2Rust} produces completed cost traces for five programs: \texttt{c4}, \texttt{yank}, \texttt{sds}, \texttt{which}, and \texttt{pwd}. 
It does not complete \texttt{cat}, \texttt{shc}, or \texttt{cflow} in our evaluation. 
For \texttt{SpeC2Rust}, the largest costs occur on \texttt{cflow} (\$36.50; 10,626\,s) and \texttt{cat} (\$11.32; 4,939\,s), which also have larger translation and repair workloads.

Overall, \texttt{SpeC2Rust} reduces cost, latency, and token usage on the programs where direct comparison is possible. It also provides completed cost traces for all eight evaluated programs.

\rqanswer{RQ3}{On the five projects completed by both tools, \texttt{SpeC2Rust} reduces total cost by 66.1\%. 
It also completes all eight evaluated projects, whereas \texttt{SmartC2Rust} completes five.}

\subsection{RQ4: Ablation Study}\label{sec:rq4-ablation}

We conducted an ablation study to understand how the major components of \texttt{SpeC2Rust} contribute to the final translation performance. 
We evaluate all variants on the eight-project dataset. 
Our evaluation is based on the following two metrics:
(i) compile rate, i.e., the fraction of projects whose translated Rust implementations compile successfully, and (ii) test pass rate, i.e., the average percentage of original end-to-end tests passed by the translated Rust implementations. Table~\ref{tab:rq4-ablation} shows the results.
The full system reaches 100.0\% on both metrics.

\begin{table}[t]
\caption{RQ4 ablation results on the eight-project dataset.}
\label{tab:rq4-ablation}
\centering
\footnotesize
\setlength{\tabcolsep}{5pt}
\renewcommand{\arraystretch}{1.10}
\begin{threeparttable}
\begin{tabular}{lcc}
\toprule
\textbf{Configuration} & \textbf{Compile Rate} & \textbf{Test Pass Rate} \\
\midrule
\texttt{SpeC2Rust} & \textbf{100.0\%} & \textbf{100.0\%} \\
w/o SpecGen Agent & 87.5\% & 25.0\% \\
w/o Hierarchical Decomposition & 62.5\% & 12.3\% \\
w/o RustRepairAgent & 25.0\% & N/A \\
\bottomrule
\end{tabular}
\end{threeparttable}
\end{table}

\textbf{Layered-Feedback Repair.}
In this study, we omitted \texttt{RustRepairAgent} from \texttt{SpeC2Rust}. Table~\ref{tab:rq4-ablation} shows that the compile rate drops from 100.0\% to 25.0\%. The test result is N/A because the result sheet does not report a completed end-to-end test outcome for this variant. 
Without this stage, the system relies on the initial Rust translation and cannot iteratively incorporate compiler diagnostics. As a result, repository-level errors that require cross-file repair or build-level adjustment are difficult to resolve within the translation pipeline.

\textbf{Project-Intent Structuring.}
In this study, we omitted the \texttt{SpecGen Agent}. Table~\ref{tab:rq4-ablation} shows that the ablated system still compiles 87.5\% of the projects, but the test pass rate drops to 25.0\%.
This gap indicates that Stage~1 affects semantic correctness more than syntactic validity. 
In other words, the ablated system can still produce Rust projects that compile in most cases, but the compiled code is less likely to satisfy the original end-to-end tests. Without the specification documents generated by Stage~1, the test agent has less repository-level context during repair and cannot reliably use the intended architecture to complete missing or incorrect functionality.

\textbf{Hierarchical Decomposition.}
In this study, we removed hierarchical decomposition from the translation process. 
Table~\ref{tab:rq4-ablation} shows that the compile rate decreases to 62.5\%, and the test pass rate decreases to 12.3\%.
Compared with the \texttt{SpecGen Agent} ablation, this variant affects both compilation and test behavior more heavily. 
This result suggests that hierarchical decomposition is important for organizing project context before specification generation and code synthesis. Without it, related files and functions are not to be grouped into coherent translation units, which makes the generated specifications and repair prompts less stable.

\textbf{Other LLMs beyond GPT.}
We also ran the full \texttt{SpeC2Rust} pipeline with
\texttt{deepseek-v4-flash}. 
It successfully translated all eight projects and passed all tests, while incurring lower overall cost than our default setting.
For this model, we set \texttt{max\_tokens} to 32K, because \texttt{deepseek-v4-flash} often emits long reasoning traces and these traces are counted within the output budget.
This suggests that the effectiveness of \texttt{SpeC2Rust} is not limited to a single underlying model, although model-specific output-length characteristics should be accounted for when configuring generation budgets.

\rqanswer{RQ4}{Removing any component reduces either compile rate or test pass rate. The largest compilation drop occurs without RustRepairAgent, while the largest semantic drop appears when project structuring or decomposition is removed.}

\section{Discussion}\label{sec:discussion}

\subsection{Why SpeC2Rust Works ?}
We use \texttt{pwd} as a case study to illustrate how \texttt{SpeC2Rust} addresses the C Project Intent challenge through specification-driven generation.
In the original C implementation, the function \texttt{robust\_getcwd} walks up the directory tree toward the root and compares device and inode numbers at each level
to reconstruct the absolute path. 
This fallback logic is important when the system \texttt{getcwd} fails, for example because of \texttt{PATH\_MAX} limits or restricted permissions on parent directories. Figure~\ref{fig:pwd-casestudy} compares the \texttt{robust\_getcwd} translations produced by \texttt{SmartC2Rust} and \texttt{SpeC2Rust} from the same C source.

\subsubsection{SmartC2Rust: test-passing-oriented translation.} As shown in Figure~\ref{fig:pwd-casestudy}(a), \texttt{SmartC2Rust} replaces the original directory-walking algorithm with a simple \texttt{env::current\_dir()} call followed by path normalization.
The translated code omits \texttt{find\_dir\_entry}, inode comparison, and the upward traversal logic.
As a result, it passes through a simpler implementation rather than preserving the original algorithmic design. This reflects a broader limitation of feedback-driven repair: the LLM is encouraged to satisfy the available test suite, but not necessarily to retain the intent of the original C implementation.

\subsubsection{SpeC2Rust: intent-preserving reconstruction.} In contrast, Figure~\ref{fig:pwd-casestudy}(b) shows that \texttt{SpeC2Rust} reconstructs the full directory-walking algorithm.
The translation preserves the \texttt{while}-loop that climbs toward the root, uses \texttt{find\_dir\_entry} to resolve each directory entry by device and inode comparison, and calls \texttt{file\_name\_prepend} to assemble the path incrementally.
It also preserves the original modular organization: the main logic remains in \texttt{pwd.rs}, inode comparison is isolated in \texttt{same\_inode.rs}, and \texttt{xgetcwd} is placed in a separate module.

This example highlights the role of \texttt{SpeC2Rust}'s \textbf{Project-Intent Structuring}. 
By explicitly recovering the intended behavior of \texttt{robust\_getcwd}---directory ascent plus inode-based name resolution---the system provides generation with stable, reusable specifications rather than relying on tests alone. 
During \textbf{Spec-Guided Generation}, these specifications guide the LLM toward reconstructing the original algorithmic structure.

The \textbf{Hierarchical Module Split} further strengthens this process by assigning supporting utilities such as \texttt{same\_inode} and \texttt{xgetcwd} to separate modules. 
This allows each module to be generated with focused specification context, reducing interference from unrelated source fragments. 
Consequently, \texttt{SpeC2Rust} is able to avoid a common failure mode of baseline tools: producing implementations that pass tests but diverge from the original project intent.
\begin{figure}[t]
\centering
\begin{minipage}{\columnwidth}
\begin{lstlisting}[language=Rust,basicstyle=\ttfamily\scriptsize,numbers=left,xleftmargin=2.2em,framexleftmargin=2.2em,framesep=5pt,escapeinside={(*@}{@*)}]
fn rust_robust_getcwd() -> Result<String, String> {
    (*@\codebad{rust\_physical\_getcwd()}@*).or_else(|_| {
        let cwd = (*@\codebad{env::current\_dir()}@*)
            .map_err(|e| e.to_string())?;
        let mut normalized = PathBuf::from("/");
        for component in cwd.components() {
            if let Component::Normal(part) = component {
                normalized.push(part);
            }
        }
        Ok(normalized.to_string_lossy().into_owned())
    })
}
// (*@\codebad{NOTE: find\_dir\_entry, inode comparison,}@*)
// (*@\codebad{and directory-walking logic are all lost.}@*)
\end{lstlisting}
\centering\footnotesize
(a) \texttt{SmartC2Rust}: algorithm replaced by \texttt{env::current\_dir()} wrapper.
\end{minipage}

\vspace{4pt}

\begin{minipage}{\columnwidth}
\begin{lstlisting}[language=Rust,basicstyle=\ttfamily\scriptsize,numbers=left,xleftmargin=2.2em,framexleftmargin=2.2em,framesep=5pt,escapeinside={(*@}{@*)}]
pub fn (*@\codegood{robust\_getcwd}@*)(file_name: &mut FileName)
    -> (*@\codegood{io::Result<()>}@*) {
    let original = env::current_dir()?;
    let root = PathBuf::from("/");
    let mut dot_meta = fs::metadata(".")?;
    let root_meta = fs::metadata(&root)?;
    // Climb toward root, comparing inodes.
    (*@\codegood{while !SameInode::same(\&dot\_meta, \&root\_meta) \{}@*)
        let h = (*@\codegood{file\_name.path.bytes()}@*)
            .filter(|b| *b == b'/').count()+1;
        (*@\codegood{Self::find\_dir\_entry}@*)(&mut dot_meta,
            file_name, h)?;
    \}
    if file_name.path.is_empty() {
        (*@\codegood{Self::file\_name\_prepend(file\_name, "");}@*)
    \}
    env::set_current_dir(original)?;
    Ok(())
}
// (*@\codegood{Module structure preserved: pwd.rs,}@*)
// (*@\codegood{same\_inode.rs, xgetcwd.rs.}@*)
\end{lstlisting}
\centering\footnotesize
(b) \texttt{SpeC2Rust}: full algorithm preserved with modular decomposition.
\end{minipage}
\caption{Comparison of \texttt{robust\_getcwd} translation in \texttt{pwd}.}
\label{fig:pwd-casestudy}
\end{figure}

\subsection{Threats to Validity}
We identify three main threats to the validity of our study.

\textbf{Internal Validity.}
A primary threat is the non-deterministic nature of LLM-based translation and repair.
Different runs may produce different Rust implementations, repair trajectories, and token costs.
To reduce this threat, we run each project multiple times and report averaged results.
Another threat concerns the quality of project knowledge produced in Stage~1.
\texttt{SpeC2Rust} relies on Stage~1 to recover module boundaries, interface relations, and behavioral constraints from the input C repository~\cite{AlphaTrans,RepoCoder}.
Errors in this stage can propagate to later generation and repair steps.
For example, missing behavioral constraints or incorrect decomposition of tightly coupled functions and data structures may lead to incomplete repair context.
The ablation results in RQ4 show that both \texttt{SpecGen} and hierarchical decomposition affect end-to-end correctness~\cite{RepoBenchRust}.

\textbf{External Validity.}
Our evaluation covers eight repository-level CLI C projects, but does not include library projects.
This scope may limit the generality of our findings to other types of C systems and build configurations.
Another possible threat is benchmark contamination. 
Some evaluated projects may have appeared in the model's pretraining data.
However, our task requires repository-level reconstruction, cross-file consistency, compilation repair, and test-based validation, which reduces but does not eliminate this concern.

\textbf{Construct Validity.}
We use the original end-to-end test scripts as the main validation criterion.
However, passing the original end-to-end tests does not prove full behavioral equivalence, because untested paths may still be incorrect~\cite{CureWorseDisease,BetterTestsAPR,RustAssure}.
Remaining \texttt{clippy} warnings may also reflect weakly exercised helper code, fallback paths, or interface fragments.
Future work could combine the current repair loop with fuzzing, differential testing, or specification-level assertions~\cite{Csmith,KLEE,RustAssure,Daikon}.
% \texttt{SpeC2Rust} uses the original project tests to check repository-level behavior. 
% Passing these tests does not prove full semantic equivalence~\cite{CureWorseDisease,BetterTestsAPR,RustAssure}. 
% Untested inputs, rare error paths, platform-specific behavior, and macro-dependent branches may still expose differences. 
% Future work could combine the current repair loop with fuzzing, differential testing, or specification-level assertions~\cite{Csmith,KLEE,RustAssure,Daikon}
For cost, we report token usage and monetary cost under the model pricing used in our experiments. 
These numbers may change with provider pricing or model updates.
\section{Related Work}\label{sec:related-work}

Related work most relevant to \texttt{SpeC2Rust} falls into three lines: rule-based and safety-oriented C-to-Rust translation, program-analysis-guided and LLM-based translation, and feedback-driven repair or agent-based code generation.

\textbf{Rule-based and safety-oriented C-to-Rust translation.}
Waters~\cite{Abstraction} provided an early abstraction-oriented view of program migration.
In C-to-Rust translation, \texttt{C2Rust}~\cite{c2rust} is the canonical rule-based approach: it largely preserves the original C decomposition and offers a practical migration starting point.
However, Emre et al.~\cite{aliasing} show that preserving semantic equivalence while removing raw pointers and \texttt{unsafe} code remains difficult.
Later work pushes safety refinement further, including \texttt{GenC2Rust}~\cite{GenC2Rust}, \texttt{C2SaferRust}~\cite{c2saferrust}, and \texttt{PR2}~\cite{PR2}.
These methods improve safety over direct transpilation, but they still mainly refine C-oriented skeletons rather than reconstructing the repository around explicit Rust-side specifications and module intent.

\textbf{Program-analysis-guided and LLM-based C-to-Rust translation.}
Recent work increasingly combines static analysis, prompt engineering, and LLM generation for C-to-Rust translation.
Eniser et al.~\cite{towards} and Pan et al.~\cite{Lost} highlight common weaknesses of LLM-based Rust translation, including compilation failures, inconsistency, and translation-induced bugs.
Other approaches add stronger structural guidance, including \texttt{VERT}~\cite{VERT}, a static-analysis-driven pipeline~\cite{LLM-Driven}, \texttt{C2RustXW}~\cite{C2RustXW}, \texttt{RustMap}~\cite{rustmap}, \texttt{SpecTra}~\cite{SpecTra}, and scalable whole-project translation with LLMs~\cite{Scalable}.
In the C-to-Rust setting, \texttt{EvoC2Rust}~\cite{evoc2rust} shows that skeleton-guided generation can improve structural consistency over direct prompting.
However, most existing approaches still use structure, analysis facts, or specifications mainly as prompt-time context rather than externalizing repository-level knowledge into persistent artifacts that can be reused across generation and repair.

\textbf{Feedback-driven repair and agent-based translation.}
Another related line studies iterative correction with compiler diagnostics, test feedback, and agent-style tool use.
This includes Rust compilation repair~\cite{Fixing_Rust}, broader LLM-based automated program repair~\cite{Automated}, and agentic repair or translation frameworks such as \texttt{RepairAgent}~\cite{RepairAgent}, LLM-powered translation agents~\cite{LLM-powered-agent}, and \texttt{TRANSAGENT}~\cite{TRANSAGENT}.
\texttt{SmartC2Rust}~\cite{smartc2rust} brings this paradigm directly into repository-level C-to-Rust translation by iteratively repairing generated Rust code using compilation and semantic feedback.
However, its heavy reliance on repair can allow hallucinated pseudo-implementations that pass tests superficially.
\section{Conclusion}\label{sec:conclusion}

This paper presented \texttt{SpeC2Rust}, a specification-driven pipeline for repository-level C-to-Rust translation. 
The core idea of \texttt{SpeC2Rust} is to externalize repository-level knowledge before Rust generation, preserve it as reusable specification artifacts, and selectively reinject it during generation and repair through \emph{Project-Intent Structuring}, \emph{Spec-Guided Generation}, and \emph{Layered-Feedback Repair}. 
We evaluated \texttt{SpeC2Rust} on eight repository-level open-source C projects, and the results show that it compiles successfully and passes the original end-to-end test scripts on all eight projects, while eliminating raw-pointer declarations, raw-pointer dereferences, and unsafe lines of code in the final translated Rust repositories. 
On the five projects where direct comparison with \texttt{SmartC2Rust} is possible, \texttt{SpeC2Rust} reduces monetary cost by 66.1\%, end-to-end latency by 77.3\%, and input-token usage by 84.7\%. Future work could extend this direction to broader classes of C systems, including library projects, and combine specification-guided translation with stronger validation techniques such as fuzzing and differential testing.

\bibliographystyle{IEEEtran}
\bibliography{IEEEexample}

\end{document}